\documentclass{article}
\usepackage[utf8]{inputenc}
\usepackage[T1]{fontenc}
\usepackage[utf8]{inputenc}
\usepackage{multicol}
\usepackage{multirow}
\usepackage{graphicx}
\usepackage{geometry}
\usepackage{subcaption}
\usepackage{todonotes}
\usepackage{amsmath}
\usepackage{amsfonts}
\usepackage{amssymb}
\usepackage{floatrow}
\usepackage{hyperref}

\hypersetup{colorlinks,citecolor=black,filecolor=black,linkcolor=black, urlcolor=black}
\newcommand\BibTeX{{\rmfamily B\kern-.05em \textsc{i\kern-.025em b}\kern-.08em
T\kern-.1667em\lower.7ex\hbox{E}\kern-.125emX}}

         \def\UnderWiggleTemp{\the\catcode`\@}
         \catcode`\@=11
         \ifx\UnderWiggle@Loaded\relax
           \catcode`\@=\UnderWiggleTemp
           \endinput \else \let\UnderWiggle@Loaded=\relax \fi
         \newbox\U@BoxA
         \newbox\U@BoxB
         \newdimen\U@DimenA
         \def\U@DoUnderWiggle{
           \offinterlineskip
           \vtop{
             \hbox{\vbox{\copy0}}
             \vskip 1.2pt  
             \vbox to 0.4pt{
               \hbox to\wd0{\hss\char'176\hss}
               \vskip0pt minus 1fil
             }
             \vskip 0.4pt  
           }
         }
         \def\UnderWiggle#1{{%
           \ifmmode
             \mathchoice
               {\setbox0=\hbox{$\displaystyle #1$}\U@DoUnderWiggle}
               {\setbox0=\hbox{$\textstyle #1$}\U@DoUnderWiggle}
               {\setbox0=\hbox{$\scriptstyle #1$}\U@DoUnderWiggle}
               {\setbox0=\hbox{$\scriptscriptstyle #1$}\U@DoUnderWiggle}
           \else
             \setbox0=\hbox{#1}\U@DoUnderWiggle
           \fi
         }}
         \catcode`\@=\UnderWiggleTemp
         \newcommand{\uw}{\UnderWiggle}

\title{A  Method for Detecting Murmurous Heart Sounds based on Self-similar Properties}
\author{Dixon Vimalajeewa, Chihoon Lee, Brani Vidakovic }
\date{December 2022}

\begin{document}

\maketitle

\begin{abstract}
A heart murmur is an atypical sound produced by the flow of blood through the heart. It can be a sign of a serious heart condition, so detecting heart murmurs is critical for identifying and managing cardiovascular diseases. However, current methods for identifying murmurous heart sounds do not fully utilize the valuable insights that can be gained by exploring intrinsic properties of heart sound signals. To address this issue, this study proposes a new discriminatory set of multiscale features based on the self-similarity and complexity properties of heart sounds, as derived in the wavelet domain. Self-similarity is characterized by assessing fractal behaviors, while complexity is explored by calculating wavelet entropy. We evaluated the diagnostic performance of these proposed features for detecting murmurs using a set of standard classifiers. When applied to a publicly available heart sound dataset, our proposed wavelet-based multiscale features achieved comparable performance to existing methods with fewer features. This suggests that self-similarity and complexity properties in heart sounds could be potential biomarkers for improving the accuracy of murmur detection.
\end{abstract}

{\bf Keywords:} Self-similarity, wavelet transform, fractality, multifractality, classification, heart murmurs.

\section{Introduction}
Cardiovascular disease (CVD) is one of the leading causes of death on the globe, accounting for 16\% of total deaths from all causes \cite{R1}. Cardiac auscultation, listening to the heart with a stethoscope, is a popular cost-effective screening method that helps in identifying murmurous heart sounds potentially indicative of CVDs (\cite{R2}).  A heart murmur is an extra swooshing sound that is produced by turbulent blood flow through the heart and is discernible from heartbeat sounds. When listening to the heart, healthcare professionals consider several factors, such as volume, location, pitch, timing of the murmur, and sound changes to tell if a murmur is harmful and a sign of cardiac health problems. Thus, cardiac auscultation can filter out patients without heart disease or defects by identifying the presence of heart murmurs \cite{R3}.

Identifying and interpreting cardiac murmurs by auscultation can be challenging even for expert cardiologists \cite{Lim2021}.  There are efforts to improve heart murmur detection methods using artificial intelligence (AI) algorithms. One solution is applying deep learning-based artificial intelligence (AI) algorithms to recordings from a digital stethoscope, and this has been a popular and promising method to detect cardiac murmurs with an accuracy that is similar to that of expert cardiologists \cite{Lim2021, Varghees2017}. For instance, \cite{R8} approached murmur detection with ResNet deep convolutional neural network and achieved 76.3\% and 91.4\% for sensitivity and specificity, respectively. The study by \cite{R9} proposed an ensemble of two convolutional neural networks (CNN) achieving 89.22\% accuracy with 89.94\% sensitivity and 86.35\% specificity. A more recent study conducted by \cite{Almanifi2022} showed that transfer learning method is a more effective method as it can achieve better performance with shorter training time.

It has also been successfully demonstrated that wavelets can be used to detect heart murmurs by utilizing  time-frequency domain properties of cardiac sounds. Unlike Fourier transforms, wavelets shift the time-domain data to a space localized in both time and frequency, thus enabling the analysis of cardiac signals at different scales. More precisely, wavelet analysis involves decomposing a signal into discrete frequency components, where the frequency is the reciprocal od scale. The decomposed coefficients are then used to  unveil hidden patterns in cardiac signals. According to \cite{Cherif2010}, discrete wavelet transforms are better at filtering heart murmur sounds without affecting different sound patterns that exist in heart sound signals. The study in \cite{R10} reported that wavelet-based techniques can achieve better performance in phonocardiogram analysis, compared to time-domain techniques. Moreover, the joint use of wavelets-based methods with NN has also been a popular approach. For example, \cite{R13} adopted features acquired from discrete wavelet transforms (DWT) in Deep Neural Networks (DNN) and support vector machines (SVM), achieving strong model performance. In a more recent study, \cite{patwa2023} used wavelet scattering transforms and 1-D convolution NNs (CNNs) to detect heart murmurs from phonocardiogram recordings automatically and accurately. This method involves pre-processing steps, such as denoising, segmentation, re-labelling of noise-only segments, data normalization, and time-frequency analysis of the phonocardiogram segments using wavelet scattering transforms.

Overall, the use of deep learning-based AI algorithms have been popular for heart murmur studies due to their ability to perform well with numerous features and to overcome the challenges of feature selection. The black-box characteristic of deep learning can make it challenging to understand how the algorithm derived its result, which can be problematic for healthcare professionals who need to provide appropriate treatments to patients \cite{R14}. Also, most of these AI models for heart murmur detection rely on time domain properties in heart sounds and pay less attention to key components of the heart sound signals in frequency domain. On the other hand, wavelet transforms have also been used to extract valuable features for those AI models in order to improve the accuracy. However, most existing studies only use wavelet transform as a pre-processing tool and do not take advantage of valuable hidden properties in heart sounds, such as fractality and self-similarity that can be unlocked through wavelet analysis and could be useful for murmur detection.

In this study we present a set of novel features based solely on wavelet transforms for detecting heart murmurs. These features are based on the self-similarity and complexity of heart sound signals analyzed in the wavelet domain. Self-similarity is characterized by the presence of stochastic similarity at different scales, which can be either monofractal or multifractal. Monofractal processes can be characterized using a single irregularity index, but more complex natural processes require a distribution of irregularity indices. This study assesses self-similarity, considering both monofractal and multifractal properties in heart sound signals. The complexity of heart sound signals is assessed by computing the wavelet entropy, that is the entropy of signal in the wavelet domain at different scales.

The analysis of monofractal properties of heart sounds usually involves assessing the scaling behavior of signals at different resolutions in the wavelet domain by using wavelet spectra. This approach characterizes the level-wise decay in scale-specific average ``energies" of the wavelet coefficients obtained from the wavelet decomposition. The term ``energy" used here is an engineering term for the  magnitude of squared wavelet coefficient. The rate of this energy decay along multiresolution scales is used as a discriminatory feature to quantify the degree of regularity in heart sound. Additionally, the normalized level-wise energies are used to compute wavelet entropy, which characterizes energy disbalance in the signal. However, in this study, we use a rolling window-based approach to quantify self-similarity and entropy, which allows for exploring localized irregularity and energy disbalance in heart sound. This approach allows analyzing the variability in irregularity and entropy and is more effective than relying on a single irregularity index and entropy.

To assess multifractality in heart sound signals, a multifractal spectrum is computed in the wavelet domain and its properties are explored. The multifractal spectrum is a function that describes the distribution of fractal dimensions across different scales of the system. To construct the spectrum, local singularity strength or irregularity is calculated at each point in the signal, and the distribution of these values across scales is measured. The multifractal spectrum can provide valuable insights into the complex behavior of the signal and can help to understand what drives it. It also enables the extraction of additional features to characterize hidden signal dynamics that are not possible with the spectra computed for monofractal processes. The degree of deviation from monofractality can be determined by exploring features such as broadness, left slope, and left tangent, which will be described in the sequel.

Finally, we evaluate the potential of these multiscale features for detecting murmurs in the heart using different classifiers, including Logistic Regression (LR), K-nearest neighbor (KNN), support vector machine (SVM), and neural network (NN), by reporting their sensitivity, specificity, and classification accuracy.

The remainder of the paper is organized as follows: Section \ref{sec:data} gives an overview of the motivating study and datasets used for the analysis. The techniques used in this study, fundamentals of wavelet transform, and assessment of monofractal and multifractal properties by using wavelet transforms, are presented in Section \ref{sec:Methods}. Sections \ref{sec:DataAnalysis} and \ref{sec:Results} provide data analysis procedures and results, respectively. Section \ref{sec:Discussion} which discusses the results is followed by some concluding remarks in Section \ref{sec:Conclusion}.

\section{Motivation Study}\label{sec:data}
\subsection{Dataset}\label{sec:data1}
\begin{figure}[!t]
\centering
\begin{subfigure}{.45\linewidth}
  \centering
  \includegraphics[width=1\textwidth]{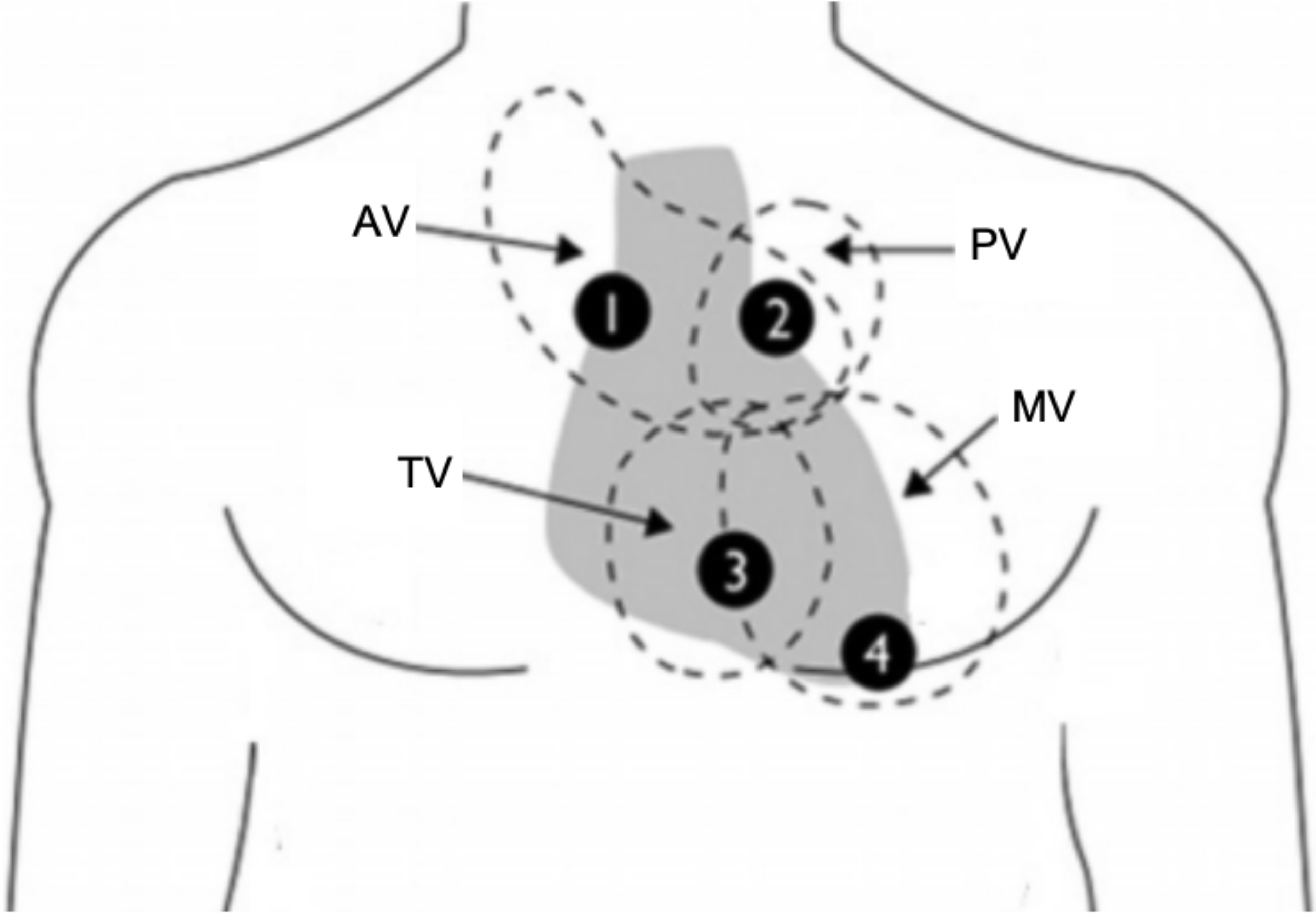}
  \caption{}
  \label{fig-11}
\end{subfigure}
\begin{subfigure}{.45\linewidth}
  \centering
  \includegraphics[width= 1\textwidth]{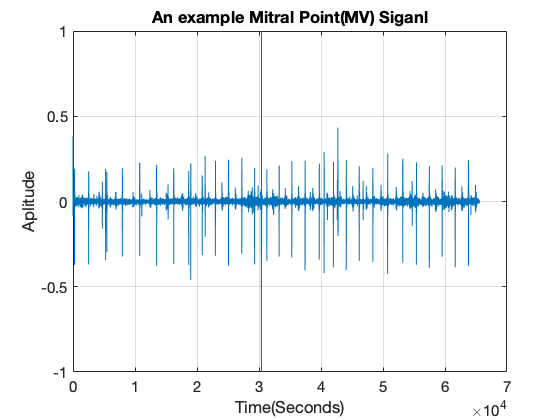}
  \caption{}
  \label{fig-12}
\end{subfigure}
\caption{(a) Locations of the heart sound recordings: \emph{Pulmonary Valve (PV), Tricuspid Valve (TV), Aortic Valve (AV)}, and \emph{Mitral Valve (MV)}; 1 = right second intercostal space; 2 = left second intercostal space; 3 = mid-left sternal border (tricuspid); 4 = fifth intercostal space, midclavicular line (the image is taken from \cite{Jorge2022}) and (b) a sample heart sound recording at \emph{Mitral Valve (MV)} point.}
\label{fig-1}
\end{figure}

This study analyzes \emph{CirCor DigiScope} dataset from \emph{Physionet 2022} \cite{Reyna2022}. The dataset was gathered in two cardiac screening campaigns on participants from rural and urban areas in Northeast Brazil. A total of 5282 heart sound recordings were collected from the main four auscultation locations of 1568 patients, 70\% of which are publicly released. Out of the total 1568 patients, 1144 (73\%) were normal subjects and 305 (19.5\%) were with heart murmurs. The four auscultation locations are \emph{Pulmonary Valve (PV), Tricuspid Valve (TV), Aortic Valve (AV)}, and \emph{Mitral Valve (MV)} (see Figure \ref{fig-1}). The ``{\it Caravan of the Heart}'' campaign, the data collector, gathered heart sounds with \emph{DigiScope collector} technology and saved as phonocardiogram (PCG) signals. These signals have a duration between 4.8 to 80.4 seconds with a mean of 22.9 and a standard deviation of 7.4 in units of a second. The sound signals were sampled at 4 kHz with 16-bit resolution and normalized to the range between -1 and 1.

The dataset was manually annotated  by an expert annotator. Signals of 119 (7.6\%) patients were inconclusive for murmur detection due to poor audio quality. The patients were aged between 0.1 and 356.1 months with a mean of 73.4 and a standard deviation of 50.3 in units of a month. Additionally, socio-demographic and clinical variables, such as age,  gender, weight, and height along with murmur location, pregnancy statures, and type of murmur (systolic or diastolic) of each patient were also compiled into a spreadsheet, where all entries were examined for data quality assessment.

\subsection{Related Works}\label{sec:RelatedWork}
In addition to the \emph{CirCor DigiScope} dataset, there are several publicly available heart sound datasets for the detection of different cardiovascular diseases, including murmurs, as described in \cite{Jorge2022}. Two of the datasets that have been used heavily for heart murmur detection are \emph{PASCAL CHSC 2011} and \emph{PhysioNet2016}.

These datasets have led to the development of several computer-aided methods for detecting heart murmurs. One popular method involves the use of deep learning (DL)-based AI algorithms \cite{Gregory2021, degroff2001}. Using the \emph{CirCor DigiScope} dataset, \cite{R15}  attained 82\% accuracy with Dual Bayesian ResNet and XGBoost integration. Despite the fact that a weighted NN-based procedure was used in this study to overcome issues of data imbalance between cases and controls, there was a significant trade-off between model sensitivity and specificity. \cite{R16} also used the same dataset to propose an ensemble classifier  in order to detect heart murmur and clinical abnormality. The proposed classifier obtained 73.7\% weighted accuracy in murmur detection. It is particularly challenging to detect murmurs in children. This is because most heart murmurs are innocent musical murmurs which should be distinguished from pathological murmurs. However, the study by \cite{degroff2001} used a smaller dataset which consisting of 37 cases and 32 controls to propose an ANN-based method and showed that it could achieve 100\% sensitivity and specificity in detecting murmur in children.  The paper \cite{Almanifi2022} uses transfer learning to detect heart murmurs from PCG recordings. Three DL models (VGG16, VGG19, and ResNet50) were evaluated on the \emph{PASCAL CHSC 2011} dataset, with ResNet50 performing the best (87.65\% accuracy).

Besides the deep learning and AI methods, attempts have been made to use wavelet transforms to develop methods for the detection of murmur. These techniques involve using wavelet transforms to decompose heart sound signals and using wavelet domain features to discriminate between normal and murmurous heart sounds. One such method involves measuring the simplicity of the wavelet-decomposed heart sound signal and adaptively thresholding it to discriminate between normal heart sounds and murmurs. This method has been tested with stenosis and regurgitation heart sounds and achieved 89.10\% sensitivity and 95.50\% specificity \cite{Varghees2017}.  Similarly, a wavelet transforms (WT) and NN based  proposed in \cite{patwa2023}  used \emph{CirCor Digiscope 2022} dataset and \emph{PCG 2016} dataset for validating the model. The study by \cite{chen2012} also used wavelets for extracting  features to detect murmur, especially in children. The ANN and SVM classifiers formed by using the extracted features showed 89\% sensitivity, 85\% specificity, and 87\% accuracy in distinguishing musical murmurs from pathological murmurs in children.

Overall, these murmur detection techniques perform well in terms of accuracy. However, there are some limitations. For instance, while neural network (NN) and deep learning-based methods can achieve state-of-the-art performance, they have limited interpretability and are computationally expensive. On the other hand, wavelet-based methods are computationally simple and can help in development of more interpretable and scalable methodologies \cite{Dixon2023_1, Dixon2023_2}. However, existing murmur detection methods do not fully utilize the potential of wavelets. That is, in the published literature, wavelets are primarily used as a data preprocessing technique and valuable features that could be extracted via wavelet transforms, such as self-similarity, fractality, compressibility, and long memory, have not been taken into account. These features have shown great potential in disease diagnostics. Therefore, our goal in this study is to create a more interpretable model while achieving comparable performance to existing models developed with the \emph{CirCor DigiScope} dataset.

\section{Methodology}\label{sec:Methods}
This section introduces the techniques used to extract multiscale features by assessing  self-similar properties of heart sound signals in the wavelet domain. First, we provide a brief overview of wavelet transforms. Then, we describe how the self-similarity is assessed by using the monofractal and multifractal spectra. Appendices A-E provide technical details about these techniques.

\subsection{Discrete Wavelet Transform}\label{sec:wt}
Wavelet transforms (WTs) are widely used tools in signal processing. When applied to a signal, they decompose the signal into a set of localized contributions in both the time and frequency domains, producing a hierarchical representation. This representation allows for simultaneous analysis at different resolutions or scales, making it possible to explore signal properties that may not be apparent in the domain of original data (acquisition domain).

The discrete wavelet transform (DWT), a popular version of wavelet transforms, has emerged as an important tool for analyzing complex data signals in application domains where discrete data are analyzed. DWTs are linear transforms that can be represented by orthogonal matrices. The computational cost of performing the DWT in matrix form increases with increasing signal length. A fast algorithm based on filtering proposed by Mallat is used for computational efficiency \cite{Mallat1989}. The DWT with this algorithm is obtained by performing a series of successive convolutions that involve a wavelet-specific low-pass filter and its mirror counterpart, high-pass filter. These repeated convolutions using the two filters accompanied with the operation of decimation (keeping every second coefficient of the convolution) generate a multiresolution representation of a signal, consisting of a smooth approximation and a hierarchy of detail coefficients at different resolutions (or scale indexes) and different locations within the same resolution. Technical details for computing these coefficients can be found in Appendix A. These coefficients describe the signal at different resolutions (scales) and locations, and are useful to characterize different multi-scale dynamics in the signal.

\subsection{Extraction of Multiscale Features}\label{sec:fract}
Fractality is an intriguing property observed in processes that exhibit self-similar behavior, which is defined by the presence of stochastic similarity  at different scales. Natural processes typically exhibit fractality in one of two forms: monofractal or multifractal, which can be summarized by the Hurst exponent (also called the irregularity index), denoted as $H$. While monofractal processes can be characterized using a single irregularity index, many natural processes are more complex and require either a "time-changing" Hurst exponent (multifractional Brownian motion) or even a distribution of irregularity indices at local times in the signal (multifractal processes). A growing body of research demonstrates that wavelets-based methods are particularly effective for analyzing fractality \cite{Goncalves1998}.

\subsection{Assessment of Monofractal Properties}\label{sec:self-similar}
Monofractality is the property of a signal where the scaling properties remain regular across the scales. This is often observed in signals that have a simple and regular structure, and in such cases, the scaling properties are uniform across all scales, and the system exhibits a single fractal dimension. In general, analysis of monofractality is useful in the study of complex systems because it provides a reference point for comparison with multifractal systems.

\begin{figure}[!t]
\centering
\begin{subfigure}{.55\linewidth}
  \centering
  \includegraphics[width=1\textwidth]{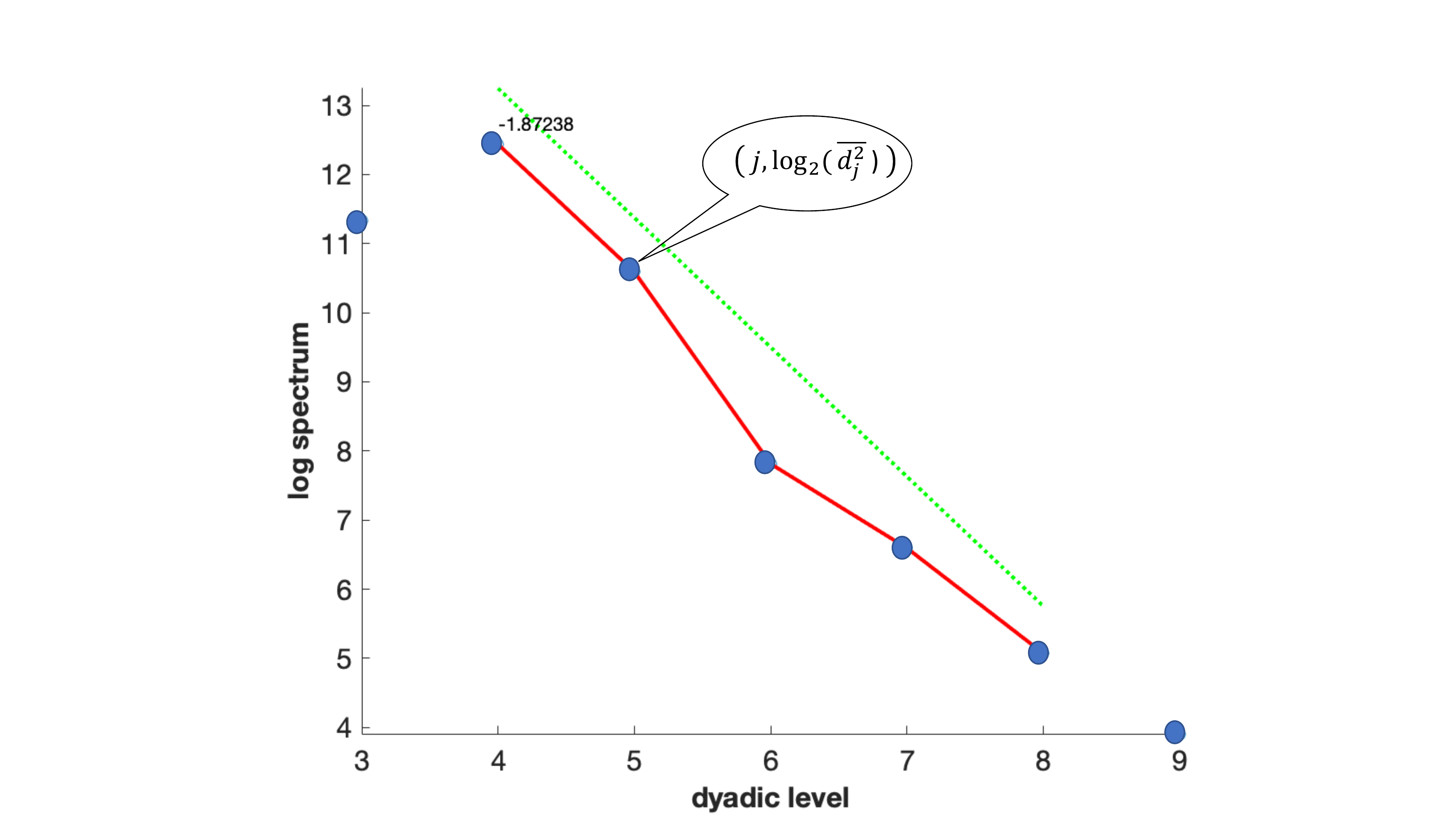}
  \caption{}
  \label{Mofrac}
\end{subfigure}
\begin{subfigure}{.425\linewidth}
  \centering
  \includegraphics[width= 1\textwidth]{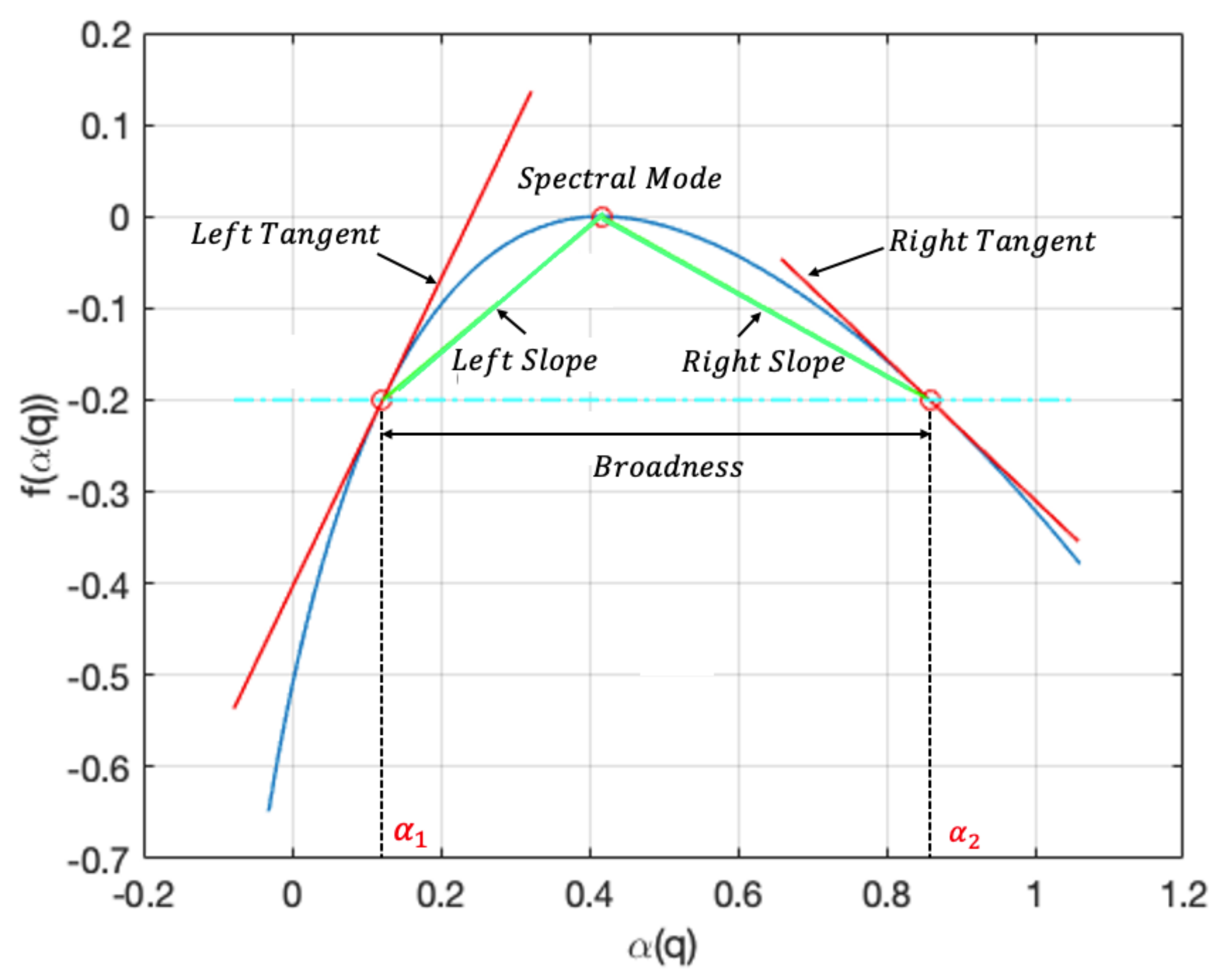}
  \caption{}
  \label{Mfrac}
\end{subfigure}
\caption{(a) A sample monofractal wavelet spectra. Slope of the wavelet spectra is estimated by fitting a straight line (green dashed) on the log energy of the wavelet coefficients (black) within the scale index ranging from 1 to 10 (red line). The coordinate of the point at the level j is $\log_2 \bar{d_j^2}$, where $d_j$ is the wavelet coefficients at the scale index j (see Appendix A for more details on wavelet coefficients $d_j$) and (b) Multifractal spectrum and its geometric descriptors; x-axis represents the irregularity index ($H$), $\alpha(q)$ and y-axis represents values proportional to the relative frequency of $H$, $f(\alpha(q))$.}
\label{fig-111}
\end{figure}

Wavelet transform based methods have been proven to be suitable for modeling monofractal processes. More specifically, wavelet spectrum  is usually used to characterize scaling behavior of the signal. The wavelet spectrum is formed by taking the log average of squared detail wavelet coefficients, which are also referred to as log energies, at different scales. Signals which possess scaling behavior (or self-similarity) exhibit a particular behavior in their wavelet spectra such that the log energies decay linearly as resolution decreases or scale increases (see Figure \ref{Mofrac}). The rate of this energy decay (slope), which is usually estimated by regressing the log  energies on the scale indices, characterizes the degree of regularity. This is usually expressed by the Hurst exponent, $H$ such that $H = (slope +1)/2$. Larger slopes ($> -2$) indicate a higher degree of persistence, and smaller slopes ($< -2$) indicate a higher degree antipersistency and intermittency. The standard Brownian motion is a boundary case with a log spectrum with slope -2. Technical details on monofractality and computing the standard wavelet spectra can be found in Appendices B and C.

\subsubsection{ Wavelet Entropy}
Wavelet entropy (WE) is commonly utilized as a natural indicator of the compressibility or complexity of signals. The complexity of a signal, which is generated by a random process, can vary depending on the underlying mechanism of randomness. For instance, if the signal is generated through a Gaussian iid process, it possesses the maximum entropy among all random generation mechanisms with a fixed mean and finite variance, indicating that the signal is not easily compressible. Furthermore, a standard Gaussian signal in the time domain corresponds to a standard Gaussian signal in the wavelet domain, where all level-wise wavelet entropies are theoretically identical. However, when a signal's generating mechanism deviates from a Gaussian distribution, the level-wise entropies in its wavelet representation can provide valuable information. In extreme cases, an ordered signal, such as a sinusoidal signal, exhibits a narrow peak in the wavelet domain, resulting in low entropy. For more in-depth information regarding the wavelet entropy of signals, please refer to \cite{Osvaldo2001}. In the present study, the normalized Shannon entropy of detail wavelet coefficients is used as the WE. Technical details on computing WE can be found in Appendix D.

\subsection{Assessment of Multifractal Properties}\label{sec:Multifractal}
Unlike monofractal or locally monofravtal signals, multifractality of a signal means that its structure has rich scaling properties even locally.
The multifractal spectrum is frequently employed as a powerful tool for examining multifractal properties. It represents the relative degree of richness of various irregularity indices. More precisely, the multifractal spectrum is constructed by calculating the local singularity strength or the Hurst exponent at each point in the signal, and then measuring the distribution of these values across different scales. Technical details on computing multifractal spectrum in the wavelet domain can be found in Appendix E. Readers can find additional information on definition of multifractal spectra using discrete wavelet transforms in \cite{Goncalves1998}

Rather than considering multifractal spectra as functions, they can be well summarized using a set of meaningful descriptors. Multifractal spectrum is a concave function and it can generally be described by three summaries, namely \emph{spectral mode (SM)}, \emph{broadness (B)}, and \emph{left (or right) tangent (LT/RT)} (see Figure \ref{Mfrac}). The \emph{spectral mode} is the most frequent scaling index; in the case of monofractal signals, the spectral mode coincides with the Hurst exponent ($H$). The \emph{broadness}(or bandwidth) is a more intricate descriptor of the multifractal spectrum and is computed as, $B = |\alpha_1 - \alpha_2|$, is the distance between $\alpha_1$ and $\alpha_2$, where $f(\alpha_1) = f(\alpha_2) = a$, for a selected value $a$, in practice usually $-0.2.$ Larger broadness reflects the presence of many degrees of scaling parameters in the signal and is a sign of deviation from the monofractality. The \emph{left tangent} is the slope of the tangent to the spectrum at the point $(\alpha_1, f(\alpha_1))$, and similarly, the \emph{right tangent}. The tangent descriptors also reflect the deviation from monofractality. That is, spectra with a lower left tangent slope reflect presence of multifractality since a pure monofractal process has theoretically an infinite LT \cite{Goncalves1998}.

Overall, analyzing the multifractal spectrum provides valuable insights into the complex behaviors of signals. It allows us to characterize the self-similarity and irregularity of signals and provides information about their inhomogeneity. Additionally, the multifractal spectrum enables the computation of descriptors related to the multi-scale nature of signals that are not visible in monofractal analysis. These descriptors can help explain differences between heart sound signals obtained from subjects with murmurs and those from control subjects.


\section{Data Analysis}\label{sec:DataAnalysis}

According to the literature on heart murmur analysis, the specific location of a murmur does not significantly influence its detection. Therefore, we focused our analysis only on heart sound signals with murmurs present or absent at four different locations. The steps we followed in our data analysis are as follows:

\subsection*{Step 1: Feature Extraction}
\begin{itemize}
\item {\bf Monofractal Features (slope):} The heart murmur sound signals were divided into sub-signals of dyadic size 1024, and multiscale features spectral slope and wavelet entropy were extracted using the discrete wavelet transform (DWT) on each sub-signal. The spectral slope was computed using scale indexes from 6 to 9 because of the stability of linear relationship between wavelet energies and scale indexes in this range (see Figure \ref{fig-2}). The wavelet entropy was computed using wavelet energies at the remaining scale indexes. The minimal phase \emph{Daubechies 6} wavelet was used for feature extraction because it balances the locality of representation and the smoothness of decomposing scaling function. The \emph{average slope} over sub-signals and \emph{coefficient of variation of wavelet entropy} over subsampled signals were computed as final features.

\item {\bf Wavelet Entropy:} As with the slope, the procedure is s. To calculate Shannon's entropy, normalized wavelet energies were computed for each sub-signal, and then the coefficient of variation of the wavelet entropy was computed.

\item {\bf Multifractal Features:} Multifractal analysis was employed on the entire heart murmur signals using the same \emph{Daubechies 6} wavelet filter. The following properties were examined: \emph{spectral mode, left slope, right slope, left tangent, right tangent, left tangent point, right tangent point} and  \emph{broadness}.
\end{itemize}

Finally, a total of ten multiscale descriptors were extracted to perform classification. They include \emph{slope, entropy, spectral mode, left slope, right slope, left tangent, right tangent, left tangent point, right tangent point} and  \emph{broadness}.

\subsection*{Step 2: Statistical Analysis}
In this study, the \emph{Wilcoxon rank sum} test was used to determine whether the features extracted from the heart sound signals differed significantly between the normal and murmurous groups and whether these differences could be used to classify heart sound signals as either having murmur present or murmur absent.
This test computes the sum of the ranks of the observations in one of the groups, and compares it to the sum of the ranks of the observations in the other group. The test statistic is the smaller of these two sums, and its significance is determined by comparing it to the distribution of the test statistic under the null hypothesis that the two groups have the same distribution.

\subsection*{Step 3: Classification Models}
The experiment involved the evaluation of four commonly used classification algorithms: logistic regression (LR), support vector machine (SVM), k-nearest neighbors (KNN), and neural network (NN). In the NN model, the input layer consists of seven nodes, while the output layer has a single node. The hidden layers are made of nodes  10, 8, 4, and 2, respectively.  While the rectified linear unit activation function is used in all other layers, the output layer uses sigmoid activation.

\subsection*{Step 4: Performance Evaluation}
Classification models were trained on 80\% of the rows from each feature matrix, with the remaining rows used for testing. Additionally, the training dataset was further subdivided as 80\% for training and the rest for validation. The imbalance in number of murmurous and normal individuals could increase a bias in the performance evaluation. To minimize the impact of data imbalance and ensure a fair assessment, two performance evaluation methods were employed as follows:

\begin{itemize}
\item {\bf Balanced Class Classification:} Equal number of samples were selected from normal and murmurous groups prior to splitting the data for training, validation and testing. That is, 616 subjects were randomly selected from the 2547 heart murmur absent cases to match the number of cases.  The process of data selection, splitting and model performance evaluation was repeated 100 times for each classifier, and the reported performance measures were averaged over these repetitions.

\item {\bf Weighted Class Classification:} The classifiers, LR, SVM, and NN were trained with adjusted weights. Using a grid search hyperparameter optimization method, optimal class weights for each classifier were determined by taking the class weights that generated the maximum Area-Under-Curve (AUC) of the model for validation data.

\end{itemize}

\subsubsection*{\bf Hyperparameter Optimization:} Grid search cross-validation technique was used to optimize the hyperparameters of classification models. This technique involves defining a grid of hyperparameter values and then training and evaluating a model for each combination of hyperparameters. Cross-validation is used to evaluate the performance of each model by splitting the data into training and validation sets multiple times and calculating the average validation score. In this study, 5-fold cross-validations were employed and the selection of optimal hyperparameters is based on the highest score in Area-Under-Curve (AUC) of Receiver Operating Characteristic (ROC) curve. Grid search cross-validation helps to ensure that the selected hyperparameters generalize well to unseen data and can improve the performance of a model.

The classifier performance was evaluated using sensitivity, specificity, AUC and overall correct classification rate (accuracy).

\begin{figure}[t!]
    \centering
    \includegraphics[width=.33\textwidth]{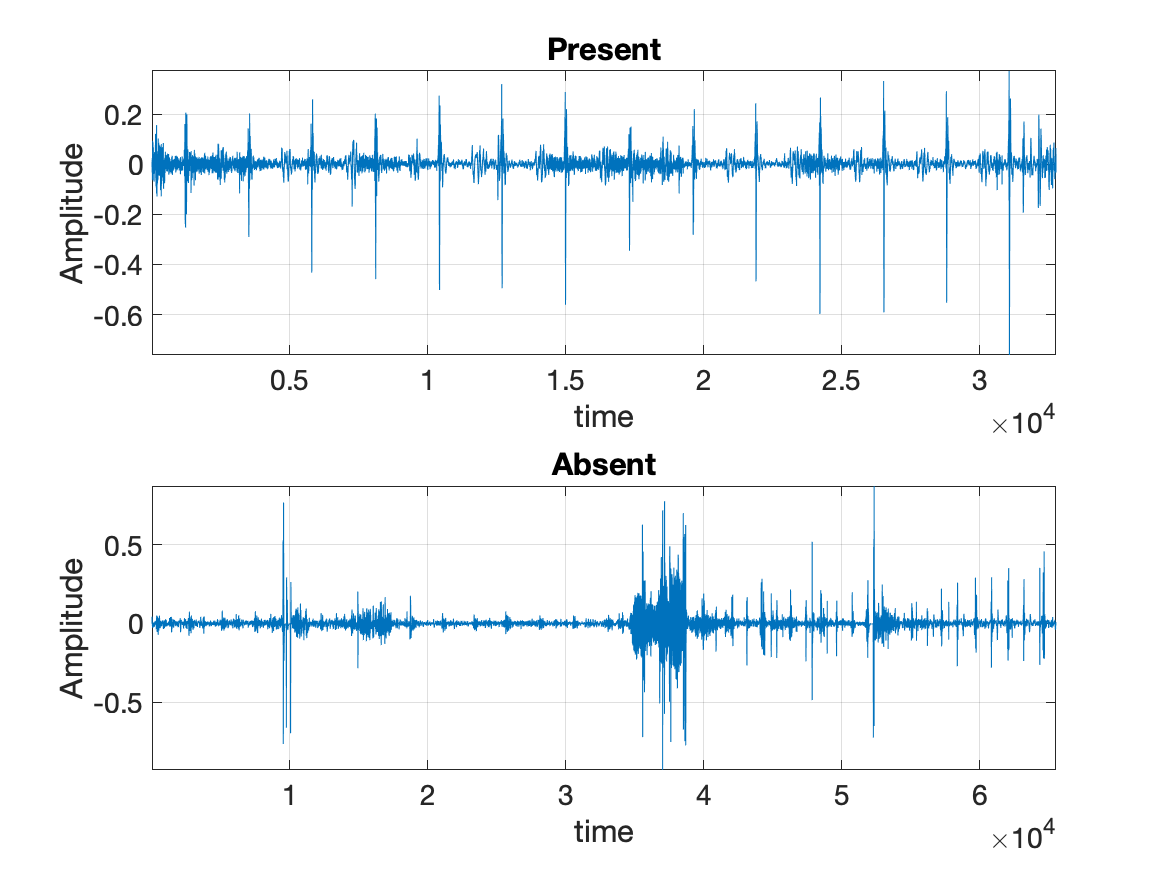}
    \includegraphics[width=.33\textwidth]{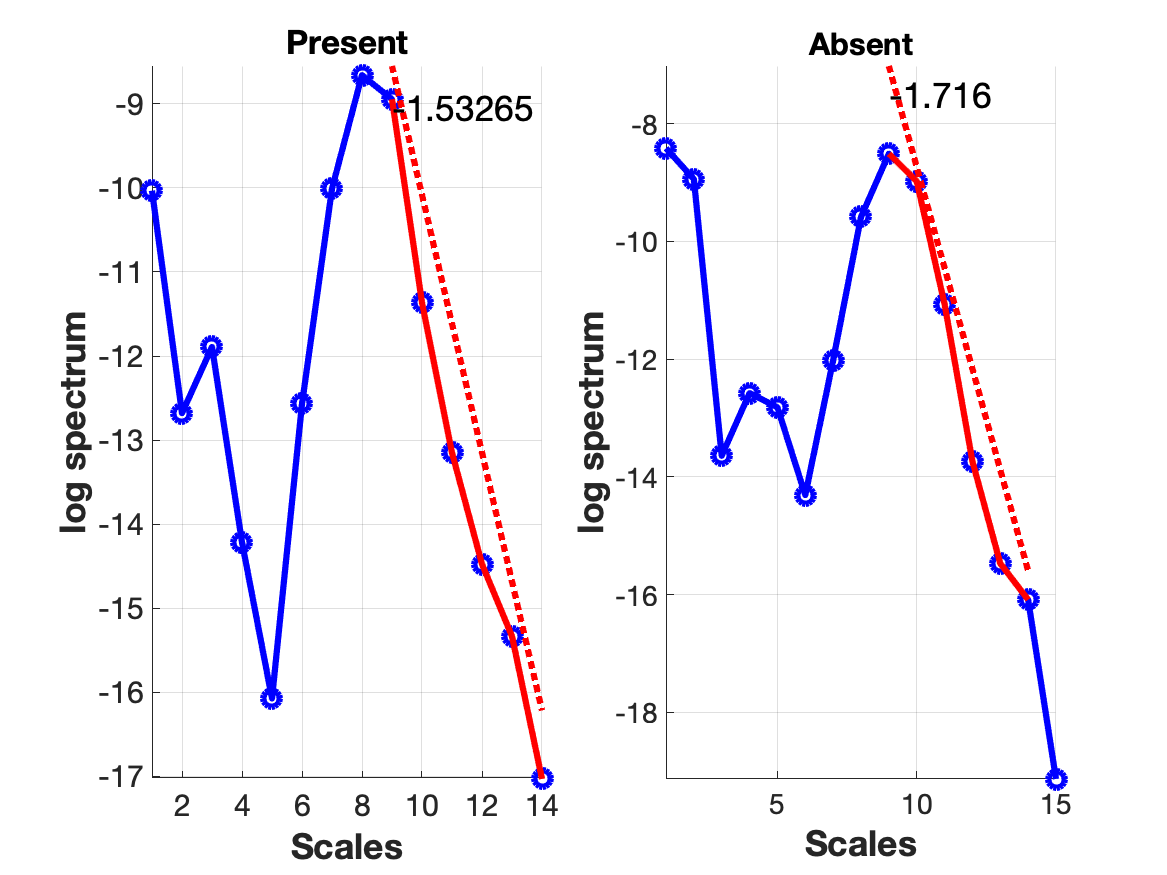}
    \includegraphics[width=.32\textwidth]{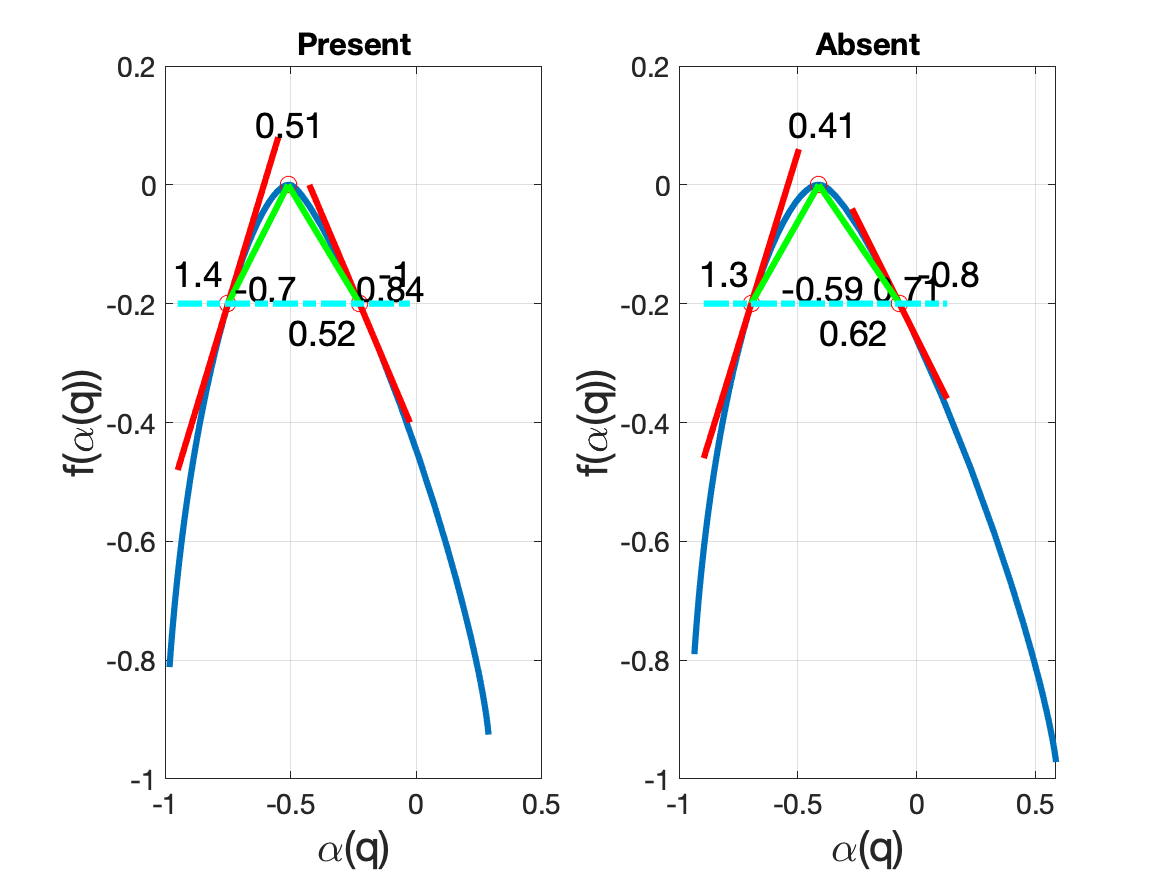}
  \caption{A sample heart sound signal with and without murmur and corresponding monofractal and multifractal spectra. For murmurs present and absent, the slopes are -1.53 and -1.72, respectively. By using the entire signal, multifractal spectra are computed, and multifractal properties are displayed as spectral modes (0.51, 0.41), left slopes (0.83, 0.71), right slopes (-0.70, -0.59), broadness (0.52, 0.62), left tangents (1.4, 1.3) and right tangents (1.00, -0.80), left tangent points (-0.75, -0.69), and right tangent points (-0.26, -0.07).}
  \label{fig-2}
\end{figure}


\section{Results}\label{sec:Results}
This section presents discriminatory behavior of the proposed features and then their potential in detecting murmurs of the heart sounds.

\subsection{Self-similar Nature in Heart Sound}
To gain a basic understanding of the self-similar nature of heart sound signals, we selected a sample of normal and murmurous heart sound signals and analyzed their monofractal and multifractal spectra. The results are shown in Figure \ref{fig-2}, which displays the differences between the fractal properties of the two types of signals. The monofractal spectra show that the spectral slope for the signal with a murmur present is -1.53, while it is -1.72 for the signal without a murmur. These values were computed using wavelet energies from scale 9 to 14. That is, the murmurous heart sound indicates a higher degree of irregularity compared to the normal heart sound. Similarly, the multifractal spectra reveal eight different properties: spectral modes (0.51, 0.41), left slopes (0.83, 0.71), right slopes (-0.70, -0.59), broadness (0.52, 0.62), left tangents (1.4, 1.3), right tangents (1.00, -0.80), left tangent points (-0.75, -0.69), and right tangent points (-0.26, -0.07). These values indicate different levels of differences in multiscale properties between the normal and murmurous heart sound signals.

\subsection{Discriminatory level of the Multiscale Features}
\begin{figure}[t]
\centering
\includegraphics[width=1\textwidth]{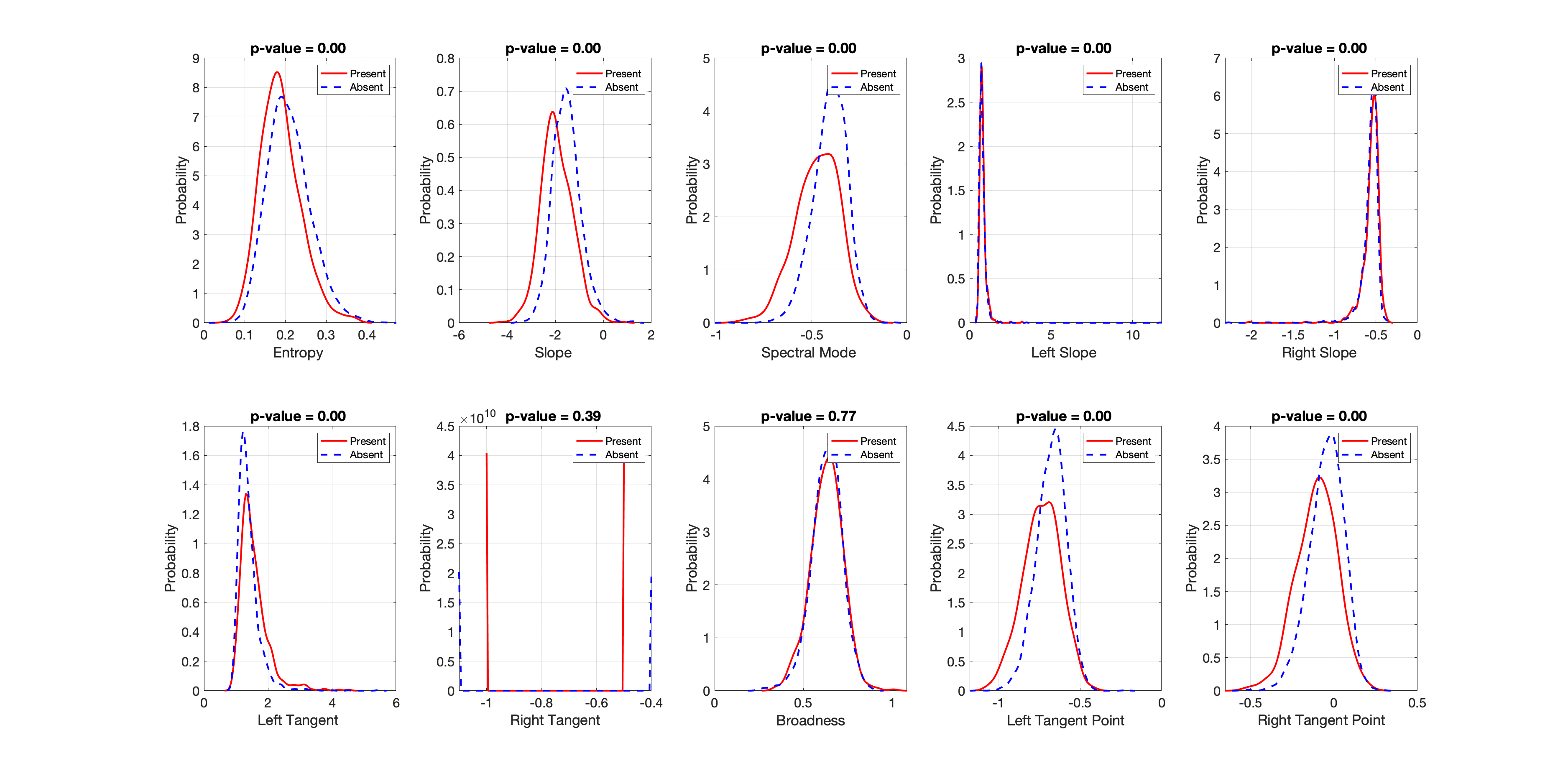}
\caption{The distribution of features proposed multiscale features.}
\label{feature_distn}
\end{figure}

Figure \ref{feature_distn} shows the distribution of individual features associated with murmurs present and absent. The results of the statistical analysis using the \emph{Wilcoxon rank sum} test indicate that the medians of most features show a significant difference between the normal and murmurous groups (with a p-value $ < 0.05$), except for the Right Tangent and Broadness in the multifractal spectra-based features. Overall, all the feature distributions exhibit varying levels of discriminatory behaviors between normal and murmurous heart sounds.

It is important to note that the statistical significance of the differences in feature medians between cases and controls does not necessarily guarantee that these features will perform well in classifying heart murmur signals. The statistical analysis serves as a preliminary step to identify potentially informative features, which can then be used to train and test machine learning models. Thus, this analysis is exploratory, designed to identify candidate features for the task of classification.

\subsection{Murmur Detection Performance with Balanced Data}\label{sec:Standard_mdl_perfom}
Table \ref{tab-1} summarizes the classification performance obtained from the balanced class classification procedure. Based on grid search, the Logistic Regression classifier was trained with \emph{saga} solver, an \emph{L1} penalty term, and an inverse regularization strength of $1 \times 10^{5}$. For the KNN classifier, we used 35 nearest neighbors and \emph{Minkowski} distance metric. The SVM classifier was based on the linear kernel with a kernel scale of 0.01 and a box constraint of 10.0.  The NN model used the ADAM optimizer with a learning rate of 0.001 and a learning rate decay of $1 \times 10^{-6}$, and binary cross as the cost function. Overall, the NN classifier achieved the highest performance with a classification accuracy of $65.70 \pm 3.99$, sensitivity of $61.61 \pm $11.15, specificity of $70.17 \pm 15.33$, and an AUC  of $0.712 \pm 0.053$. This was followed by LR, SVM, and KNN classifiers.

The  performance summarized in Table \ref{tab-2} is obtained from the weighted classifiers implemented using LR, SVM, and NN. The KNN model is not included as it is not able to train with adjusted class weights.  By utilizing the grid search with the hyperparameters and the class weights, the optimal settings for each of these classifiers were determined. The LR classifier used the \emph{saga} solver with \emph{L1} penalty term, an inverse regularization strength of 0.1, and the class weights of 1 for murmur absence and 4.5 for murmur presence. The SVM classifier used the \emph{RBF} kernel with a kernel scale of 0.1, a box constraint of 100.0, and class weights of 1 for murmur absence and 4 for murmur presence. These best class weights were found using grid search with 5-fold cross validation for LR and SVM models. For the NN model, the model parameters were as same as the NN model used in the repeated classifier method except for the class weights. Unlike SVM and LR, we explored ROC curve to find the optimal class weights for the NN model. After splitting training data into training and validation datasets, we trained a separate NN model on each of the eight different class weight values and computed Youden index from sensitivity and 1-specificity obtained from each NN model. The class weights associated with the Youden index are the optimal class weights. As a result, the class weights used for the NN model were 0.62 for murmur absence and 2.54 for murmur presence. The SVM classifier achieved the highest performance with a classification accuracy of 76.61\%, specificity of 82.12\%, sensitivity of 54.02\%, and an AUC of 0.741. This was followed by the NN and LR classifiers.

All in all, classification models with  adjusted class weights perform better than the models without class weights.

\begin{table}[!h]
\centering
\caption{ Murmur detection performance with the balanced normal and murmurous group.}
\begin{tabular}{|l|c  | c  |c | c |}
\hline
\multirow{2}{*}{\bf Classifier} &  {\bf Classification} & \multirow{2}{*}{\bf Sensitivity} & \multirow{2}{*}{\bf Specificity} & \multirow{2}{*}{\bf AUC} \\
 & {\bf Accuracy} & & &\\
\hline
Logistic  & \multirow{2}{*}{65.14 $\pm$ 2.71}  & \multirow{2}{*}{62.34 $\pm$ 4.31}  & \multirow{2}{*}{68.17 $\pm$ 4.47} & \multirow{2}{*}{0.715 $\pm$ 0.029} \\
 Regression (LR) &  &  & &  \\
\hline
Support Vector  &  \multirow{2}{*}{64.94 $\pm$ 2.68} & \multirow{2}{*}{58.42 $\pm$ 4.46} &  \multirow{2}{*}{71.85 $\pm$ 4.41} & \multirow{2}{*}{0.714 $\pm$ 0.029}  \\
 Machine (SVM) &  &  &  & \\
\hline
K Nearest & \multirow{2}{*}{63.56 $\pm$ 2.77}  &  \multirow{2}{*}{56.43 $\pm$ 4.73} &  \multirow{2}{*}{71.08 $\pm$ 4.54} & \multirow{2}{*}{0.700 $\pm$ 0.030}\\
 Neighbor (KNN) & &  & & \\
\hline
Neural & \multirow{2}{*}{\bf 65.70 $\pm$ 3.99}  &  \multirow{2}{*}{\bf 61.61$\pm$ 11.15}  &  \multirow{2}{*}{\bf 70.17 $\pm$ 15.33}  & \multirow{2}{*}{\bf 0.712 $\pm$ 0.053}\\
 Network (NN) & &  & &  \\
\hline
\end{tabular}
\label{tab-1}
\end{table}

\begin{table}[!h]
\centering
\caption{Murmur detection performance with adjusted class weights.}
\begin{tabular}{|l|c |c |c |c |c |c |}
\hline
\multirow{2}{*}{\bf Classifier} & \multirow{2}{*}{\bf Classification} & \multicolumn{2}{c|}{\bf F1-score} & \multirow{2}{*}{\bf Sensitivity} & \multirow{2}{*}{\bf Specificity} & \multirow{2}{*}{\bf AUC} \\
& \multirow{2}{*}{\bf Accuracy} & \multicolumn{2}{c|}{} & & & \\
\cline{3-4}
& & {\bf Absent} & {\bf Present} &  &  & \\
\hline
\multirow{2}{*}{\shortstack[l]{Logistic\\Regression (LR)}} & \multirow{2}{*}{63.98} & \multirow{2}{*}{74.03} & \multirow{2}{*}{44.62} & \multirow{2}{*}{68.50} & \multirow{2}{*}{65.22} & \multirow{2}{*}{0.726}\\
  &  &  &  & & & \\
\hline
\multirow{2}{*}{\shortstack[l]{Support Vector\\ Machine (SVM)}}  &  \multirow{2}{*}{\bf 76.61 } & \multirow{2}{*}{\bf 84.96} &  \multirow{2}{*}{\bf 47.52 } & \multirow{2}{*}{\bf 54.03} & \multirow{2}{*}{\bf 82.12} & \multirow{2}{*}{\bf 0.741} \\
  &  &  &  & & & \\
\hline
\multirow{2}{*}{\shortstack[l]{Neural \\ Network(NN)}} & \multirow{2}{*}{73.78} & \multirow{2}{*}{82.60} &  \multirow{2}{*}{46.79} & \multirow{2}{*}{57.48} & \multirow{2}{*}{77.87} & \multirow{2}{*}{0.719} \\
  &  &  &  & & & \\
\hline
\end{tabular}
\label{tab-2}
\end{table}
\section{Discussion}\label{sec:Discussion}
The main focus of this study is to propose a new set of multiscale features based on self-similarity and complexity properties of heart sounds in the wavelet domain and evaluate their effectiveness in detecting heart murmurs. The study also compares the proposed approach with previously proposed methods using the same dataset, demonstrating its potential to enhance the detection of murmurs.

This study employs wavelet analysis to extract features from heart sound signals in the frequency domain, offering a fresh perspective compared to traditional time-domain analysis methods. The proposed methodology demonstrates that the multiscale features extracted contribute to enhanced detection of heart murmurs compared to existing methods that primarily involve the segmentation of S1 and S2 sounds. This is because the use of multiscale features allows for the capture of unique and hidden dynamics in heart sound signals that cannot be revealed through conventional feature engineering methods typically used in statistical analyses. Additionally, discrete wavelet transforms effectively filter heart murmurs while preserving S1 and S2 sound patterns \cite{Cherif2010}. As a result, the feature distributions shown in Figure \ref{feature_distn} indicate that normal and murmurous heart sounds differ in terms of their discriminatory levels of multiscale properties. Heartbeat of healthy individuals indicates higher entropy compared to the individuals with murmurs. That is, normal heart sound exhibits a more complexity compared to murmurous sounds. The lower entropy of  murmurous heart sounds could be due to the increased energy imbalance caused by deviations from normal sound patterns, resulting in a higher concentration of signal energy in abnormal sounds. The slope in murmurous heart signals also supports the hypothesis that these signals have more ordered sound patterns compared to normal heart sounds. Furthermore, the observation that heart sounds with murmurs often possess multifractal properties is consistent with previous studies on multifractality in cardiovascular disease analysis \cite{Ivanov1999, Rufiner2018, Li2018}. Overall, these findings provide compelling evidence for the potential use of the proposed features as biomarkers for identifying individuals with murmurs, which could aid in the diagnosis and management of cardiovascular diseases.

\begin{table*}[!t]
\centering
\caption{ Performance comparison with the existing studies that used the same CirCor dataset (MPL - multilayer perception, LSTM - long-short term memory).}
\begin{tabular}{|l|l|l|c|c|c|}
\hline
{\bf Study} &  {\bf \shortstack[l]{Feature \\Domain}} & {\bf \shortstack[l]{Feature\\ Extraction}} & {\bf \shortstack[l]{Number of\\ Features}}  & {\bf \shortstack[l]{Classification\\ Accuracy}} & {\bf Classifier}\\
\hline
\multirow{3}{*}{\cite{Jalali2022}} & Time and  & Standard statistical measures    & \multirow{3}{*}{151} & \multirow{3}{*}{57.69} & \multirow{3}{*}{MLP}   \\
&Frequency  & derived from original, Fourier   && & \\
& & transformed and WT signals &&& \\
\hline
 \multirow{3}{*}{ \cite{Monteiro2022}} & Frequency & Homomorphic, Hilbert,  & \multirow{3}{*}{200} & \multirow{3}{*}{75.1} & \multirow{3}{*}{LSTM}  \\
 & & power spectral density,  &  & &  \\
 && and wavelet envelopes && &\\
\hline
\cite{ballas2022} & Time &  Data
augmentation & - & 73.7 & CNN \\
\hline
\multirow{2}{*}{ \cite{Fuadah2023}} &  Frequency  & Mel frequency,  & \multirow{2}{*}{42} & \multirow{2}{*}{76.31} &  \multirow{2}{*}{KNN}\\
& & cepstrum coefficients & & &\\
 \hline
\multirow{2}{*}{Prposed Method} & \multirow{2}{*}{Frequency} & Self-similarity & \multirow{2}{*}{8} & \multirow{2}{*}{76.61} & \multirow{2}{*}{SVM} \\
& & and wavelet entropy &  &&\\
\hline
\end{tabular}
\label{tab-3}
\end{table*}

This approach has the potential to uncover and utilize information about heart sounds that may not be easily observable in the time domain. As a result, the proposed methodology offers several advantages. One of the most significant advantages is that the classification models are simple in terms of the number of features used while still achieving performance comparable to existing methods based on the same dataset, which commonly use deep learning models. Another key advantage is that the methodology requires minimal pre-processing, unlike existing techniques that heavily rely on pre-processing heart sound signals. Since there is no standard pre-processing procedure, differences in pre-processing could limit the generalizability and reproducibility of valuable features. The proposed methodology addresses these challenges by requiring only minimal pre-processing, thus enhancing the generalizability and reproducibility of the findings. Moreover, the proposed method can achieve comparable performance to the existing method with relatively simple classifiers. Table \ref{tab-3} compares the feature extraction domain, feature extraction method, number of features and performance of our proposed method with previous studies that used the same \emph{CirCor} dataset. It can be seen that existing studies have used characteristics of heart sounds in the time, frequency and time-frequency domains to build machine learning models. When comparing the model complexity in terms of the number of features, the classifiers proposed in this study are relatively much simpler while still achieving comparable (or even better) performance.

The proposed methodology presents certain challenges that may need to be addressed in future studies. For example, as shown in Figure \ref{fig-2}, the scale index range used to calculate slope values was restricted from 9 to 14, which may not be suitable for all signals and could result in slope values outside the theoretically expected range of -3 to -1 (as illustrated in Figure \ref{feature_distn}). Similarly, selecting an appropriate range of moment $q$ that is suitable for all signals when calculating multifractal spectra also presented challenges. The range $q = [3, 12]$ was applicable to over 90\% of signals, but for some signals, it resulted in numerical instability. As a result, a shorter range of 3 to 10 was used for which the range $q = [3, 12]$ did not work. Consequently, the generalizability of the proposed procedure is limited and further investigation is needed to identify more generalizable values for these parameters. Additionally, the study used the discrete wavelet transform (DWT), which requires the signal length to be a power of two. Therefore, the maximum possible signal length that satisfies this condition was selected, resulting in some information loss. This can be overcome by replacing DWT with non-decimated wavelet transform as it allows performing wavelet transform on a signal of any length. Furthermore, the original dataset includes additional features such as age, gender, height and weight and it would be interesting to investigate how the proposed multiscale features are related to these features and their impact on the presence of murmurs.

When developing classification models, one of the main challenges of this study is the imbalanced data, with 2547 normal heart sound signals compared to only 616 signals with a murmur. This imbalance can lead to classifiers becoming biased towards the majority class, in this case, the normal heart sound signals. Although classification with adjusted class weights is an option and performs well as shown in this study, adjusting the cost function by heavily weighting the minority class can result in a significant trade-off between sensitivity and specificity. Overall, the weighting method can lead to an increase in false positives, which limits the practical usefulness of the proposed neural network model. Therefore, to improve the practical usefulness of the model, it would be more beneficial to focus on specific cardiac diseases rather than a general abnormality in heart sounds.

\section{Conclusions}\label{sec:Conclusion}
This study introduces a new set of multiscale features based on self-similar properties and entropy of heart sound signals. Self-similarity in data signals is assessed by considering monofractal and multifractal behaviors in the wavelet domain. The potential of accurately detecting murmur of these features is assessed by developing a neural network-based classifier and standard classifiers. The application of proposed features on heart sound signals showed the ability to detect murmurous heat sound with an accuracy of over 76.01\%. Compared to existing methods, the proposed approach is simpler but achieves comparable results. Therefore, the multiscale features could serve as potential biomarkers for automated heart murmur detection.

In the spirit of reproducible research, the software used in this paper is posted on
\url{https://github.com/chihoon1/Murmur.git}.
\bibliographystyle{plain} 
\bibliography{sample}

\begin{thebibliography}{10}

\bibitem{Almanifi2022}
Omair Rashed~Abdulwareth Almanifi, Ahmad~Fakhri {Ab Nasir}, Mohd~Azraai {Mohd
  Razman}, Rabiu~Muazu Musa, and Anwar {P.P. Abdul Majeed}.
\newblock Heartbeat murmurs detection in phonocardiogram recordings via
  transfer learning.
\newblock {\em Alexandria Engineering Journal}, 61(12):10995--11002, 2022.

\bibitem{ballas2022}
Aristotelis Ballas, Vasileios Papapanagiotou, Anastasios Delopoulos, and
  Christos Diou.
\newblock Listen2yourheart: A self-supervised approach for detecting murmur in
  heart-beat sounds, 2022.

\bibitem{chen2012}
Yuerong Chen, Shengyong Wang, Chia-Hsuan Shen, and Fred~K Choy.
\newblock Intelligent identification of childhood musical murmurs.
\newblock {\em Journal of Healthcare Engineering}, 3(1):125--139, 2012.

\bibitem{Cherif2010}
L.~Hamza Cherif, S.M. Debbal, and F.~Bereksi-Reguig.
\newblock Choice of the wavelet analyzing in the phonocardiogram signal
  analysis using the discrete and the packet wavelet transform.
\newblock {\em Expert Systems with Applications}, 37(2):913--918, 2010.

\bibitem{R8}
John~S Chorba, Avi~M Shapiro, Le~Le, John Maidens, John Prince, Steve Pham,
  Mia~M Kanzawa, Daniel~N Barbosa, Caroline Currie, Catherine Brooks, et~al.
\newblock Deep learning algorithm for automated cardiac murmur detection via a
  digital stethoscope platform.
\newblock {\em Journal of the American Heart Association}, 10(9):e019905, 2021.

\bibitem{degroff2001}
Curt~G DeGroff, Sanjay Bhatikar, Jean Hertzberg, Robin Shandas, Lilliam
  Valdes-Cruz, and Roop~L Mahajan.
\newblock Artificial neural network-based method of screening heart murmurs in
  children.
\newblock {\em Circulation}, 103(22):2711--2716, 2001.

\bibitem{Fuadah2023}
Yunendah~Nur Fuadah, Muhammad~Adnan Pramudito, and Ki~Moo Lim.
\newblock An optimal approach for heart sound classification using grid search
  in hyperparameter optimization of machine learning.
\newblock {\em Bioengineering}, 10(1), 2023.

\bibitem{Goncalves1998}
P.~Goncalves, R.~Riedi, and R.~Baraniuk.
\newblock A simple statistical analysis of wavelet-based multifractal spectrum
  estimation.
\newblock In {\em Conference Record of Thirty-Second Asilomar Conference on
  Signals, Systems and Computers (Cat. No.98CH36284)}, volume~1, pages 287--291
  vol.1, 1998.

\bibitem{R10}
Shahid Ismail, Imran Siddiqi, and Usman Akram.
\newblock Localization and classification of heart beats in phonocardiography
  signals—a comprehensive review.
\newblock {\em EURASIP Journal on Advances in Signal Processing},
  2018(1):1--27, 2018.

\bibitem{Ivanov1999}
Plamen~Ch. Ivanov, Lu{\'\i}s A.~Nunes Amaral, Ary~L. Goldberger, Shlomo Havlin,
  Michael~G. Rosenblum, Zbigniew~R. Struzik, and H.~Eugene Stanley.
\newblock Multifractality in human heartbeat dynamics.
\newblock {\em Nature}, 399(6735):461--465, 1999.

\bibitem{Jalali2022}
Kiarash Jalali, Mohammad~Amin Saket, and Saman Noorzadeh.
\newblock Heart murmur detection and clinical outcome prediction using
  multilayer perceptron classifier.
\newblock In {\em 2022 Computing in Cardiology (CinC)}, volume 498, pages 1--4,
  2022.

\bibitem{Li2018}
Jin Li, Chen Chen, Qin Yao, Peng Zhang, Jun Wang, Jing Hu, and Feilong Feng.
\newblock The effect of circadian rhythm on the correlation and multifractality
  of heart rate signals during exercise.
\newblock {\em Physica A: Statistical Mechanics and its Applications},
  509:1207--1213, 2018.

\bibitem{Lim2021}
Gregory~B. Lim.
\newblock Ai used to detect cardiac murmurs.
\newblock {\em Nature Reviews Cardiology}, 18(7):460--460, 2021.

\bibitem{Gregory2021}
Gregory~B. Lim.
\newblock Ai used to detect cardiac murmurs.
\newblock {\em Nature Reviews Cardiology}, 18(7):460--460, 2021.

\bibitem{Mallat1989}
S.G. Mallat.
\newblock A theory for multiresolution signal decomposition: the wavelet
  representation.
\newblock {\em IEEE Transactions on Pattern Analysis and Machine Intelligence},
  11(7):674--693, 1989.

\bibitem{R14}
Seonwoo Min, Byunghan Lee, and Sungroh Yoon.
\newblock {Deep learning in bioinformatics}.
\newblock {\em Briefings in Bioinformatics}, 18(5):851--869, 07 2016.

\bibitem{Monteiro2022}
Sofia Monteiro, Ana Fred, and Hugo~Plácido da~Silva.
\newblock Detection of heart sound murmurs and clinical outcome with
  bidirectional long short-term memory networks.
\newblock In {\em 2022 Computing in Cardiology (CinC)}, volume 498, pages 1--4,
  2022.

\bibitem{Varghees2017}
V.~Nivitha~Varghees and K.~I. Ramachandran.
\newblock Effective heart sound segmentation and murmur classification using
  empirical wavelet transform and instantaneous phase for electronic
  stethoscope.
\newblock {\em IEEE Sensors Journal}, 17(12):3861--3872, 2017.

\bibitem{R9}
Fuad Noman, Chee-Ming Ting, Sh-Hussain Salleh, and Hernando Ombao.
\newblock Short-segment heart sound classification using an ensemble of deep
  convolutional neural networks.
\newblock In {\em ICASSP 2019-2019 IEEE International Conference on Acoustics,
  Speech and Signal Processing (ICASSP)}, pages 1318--1322. IEEE, 2019.

\bibitem{Jorge2022}
Jorge Oliveira, Francesco Renna, Paulo~Dias Costa, Marcelo Nogueira, Cristina
  Oliveira, Carlos Ferreira, Alípio Jorge, Sandra Mattos, Thamine Hatem,
  Thiago Tavares, Andoni Elola, Ali~Bahrami Rad, Reza Sameni, Gari~D. Clifford,
  and Miguel~T. Coimbra.
\newblock The circor digiscope dataset: From murmur detection to murmur
  classification.
\newblock {\em IEEE Journal of Biomedical and Health Informatics},
  26(6):2524--2535, 2022.

\bibitem{patwa2023}
Ahmed Patwa, Muhammad Mahboob~Ur Rahman, and Tareq~Y. Al-Naffouri.
\newblock Heart murmur and abnormal pcg detection via wavelet scattering
  transform \& a 1d-cnn, 2023.

\bibitem{R3}
David~H. Peters, Anu Garg, Gerry Bloom, Damian~G. Walker, William~R. Brieger,
  and M.~Hafizur~Rahman.
\newblock Poverty and access to health care in developing countries.
\newblock {\em Annals of the New York Academy of Sciences}, 1136(1):161--171,
  2008.

\bibitem{Reyna2022}
Matthew~A. Reyna, Yashar Kiarashi, Andoni Elola, Jorge Oliveira, Francesco
  Renna, Annie Gu, Erick A.~Perez Alday, Nadi Sadr, Ashish Sharma, Sandra
  Mattos, Miguel~T. Coimbra, Reza Sameni, Ali~Bahrami Rad, and Gari~D.
  Clifford.
\newblock Heart murmur detection from phonocardiogram recordings: The george b.
  moody physionet challenge 2022.
\newblock {\em medRxiv}, 2022.

\bibitem{Osvaldo2001}
Osvaldo~A. Rosso, Susana Blanco, Juliana Yordanova, Vasil Kolev, Alejandra
  Figliola, Martin Schürmann, and Erol Başar.
\newblock Wavelet entropy: a new tool for analysis of short duration brain
  electrical signals.
\newblock {\em Journal of Neuroscience Methods}, 105(1):65--75, 2001.

\bibitem{Rufiner2018}
Hugo~L. Rufiner, Paolo Castiglioni, Davide Lazzeroni, Paolo Coruzzi, and Andrea
  Faini.
\newblock Multifractal-multiscale analysis of cardiovascular signals: A
  dfa-based characterization of blood pressure and heart-rate complexity by
  gender.
\newblock {\em Complexity}, 2018:4801924, 2018.

\bibitem{R16}
Sara Summerton, Danny Wood, Darcy Murphy, Oliver Redfern, Matt Benatan, Matti
  Kaisti, and David~C Wong.
\newblock Two-stage classification for detecting murmurs from phonocardiograms
  using deep and expert features.
\newblock In {\em Computing in Cardiology 2022: 49th Computing in Cardiology
  Conference}, 2022.

\bibitem{R2}
Connie~W Tsao, Aaron~W Aday, Zaid~I Almarzooq, Alvaro Alonso, Andrea~Z Beaton,
  Marcio~S Bittencourt, Amelia~K Boehme, Alfred~E Buxton, April~P Carson,
  Yvonne Commodore-Mensah, et~al.
\newblock Heart disease and stroke statistics—2022 update: a report from the
  american heart association.
\newblock {\em Circulation}, 145(8):e153--e639, 2022.

\bibitem{Dixon2023_2}
Dixon Vimalajeewa, Scott~Alan Bruce, and Brani Vidakovic.
\newblock Early detection of ovarian cancer by wavelet analysis of protein mass
  spectra.
\newblock {\em Statistics in Medicine}, 2023.

\bibitem{Dixon2023_1}
Dixon Vimalajeewa, Ethan McDonald, Scott~Alan Bruce, and Brani Vidakovic.
\newblock Wavelet-based approach for diagnosing attention deficit hyperactivity
  disorder (adhd).
\newblock {\em Scientific Reports}, 12(1):21928, 2022.

\bibitem{R15}
Ben Walker, Felix Krones, Ivan Kiskin, Guy Parsons, Terence Lyons, and Adam
  Mahdi.
\newblock Dual bayesian resnet: A deep learning approach to heart murmur
  detection.
\newblock {\em Computing in Cardiology}, 2022.

\bibitem{R1}
{WHO}.
\newblock {WHO} reveals leading causes of death and disability worldwide:
  2000-2019, 2020.

\bibitem{R13}
Yaseen, Gui-Young Son, and Soonil Kwon.
\newblock Classification of heart sound signal using multiple features.
\newblock {\em Applied Sciences}, 8(12):2344, 2018.

\end{thebibliography}

\appendix
\subsection*{Appendix A: Discrete Wavelet Transform}\label{sec:wt1}
Suppose a data signal ($Y$) is a vector of size  $N \times 1$. The DWT of $Y$, denoted as  $d$, is represented as
\begin{equation}\label{eq-12}
  d = WY,
\end{equation}
where $W$ is an orthogonal matrix of size ${N \times N}$. The elements in $W$ are determined by selecting a particular wavelet basis, such as Haar, Daubechies, and Symmlet. Although $N$ can be arbitrary, it is usually selected to be a power of two, i.e., $N = 2^J$, $J \in \mathbb{Z}^+$ for the ease of calculating the DWT.

When $N$ is large, abd calculating $d$ as in (\ref{eq-12}) becomes computationally expensive, a fast algorithm based on filtering proposed by Mallat is used for computational efficiency. With this algorithm the DWT is obtained by performing a series of successive convolutions that involve a wavelet-specific low-pass filter $h$ and its quadrature-mirror counterpart, high-pass filter $g$. These repeated convolutions using the two filters accompanied with the operation of decimation (keeping every second coefficient of the convolution) generate a multiresolution representation of a signal, consisting of a smooth approximation ($c$-coefficients) and a hierarchy of detail coefficients $d_{jk}$ at different resolutions (indexed by a scale index $j$) and different locations within the same resolution, indexed by $k$. The convolutions with filters $h$ and $g$ are repeated until a desired decomposition level $j=J_0$ is reached ($1 \leq J_0 \leq J-1 $ and $J=\log_2 N$).

Thus, the vector  $\uw{d}$ in (\ref{eq-12}) has a structure,
\begin{equation}\label{eq-12a}
  \uw{d} = (\uw{c}_{J_0}, \uw{d}_{J_0}, \dots, \uw{d}_{J-2}, \uw{d}_{J-1}),
\end{equation}
where $\uw{c}_{J_0}$ is a vector of coefficient corresponding to a smooth trend in signal, and $\uw{d}_{j}$ are detail coefficients at different resolutions $j$ where $J_0 \leq j \leq J-1.$  It is the logarithm of the variance of coefficients in $\uw{d}_j$ that is used to define wavelet spectra.

\subsection*{Appendix B: Monofractality}\label{sec:self-similar1}
A deterministic function $f(t)$ of $d$-dimensional argument $t$ is said to be self-similar, if $f(\lambda t) = \lambda^{-H}f(\lambda t)$, for some choice of the Hurst exponent $H$, and for all dilation factors $\lambda$. The notation of self-similarity has been extended to random processes. Specifically, a stochastic process $\{ X(t), t \in \mathbb{R}^d \}$ is self-similar with a scaling parameter (or Hurst exponent $H$) if, for any $\lambda \in \mathbb{R}^+$,

\begin{equation}\label{self-similar}
    X(\lambda t) \overset{d}{=} \lambda^H X(t),
\end{equation}
where the relation ``$\overset{d}{=}$'' denotes the equality in all finite-dimensional distributions.

The scaling exponent ($H$), is a statistical parameter in a well-defined statistical model when the signal follows stochastic behavior (e.g., Gaussianity and stationary increments).

Wavelet transforms based methods have been proven to be suitable for self-similar modeling processes with stationary increments. For example, models such as fractional Brownian motion (fBm) in one dimension and fractional Brownian fields (fBf) in $d$ dimensions are frequently used to explain data that scale. For a comprehensive description, see \cite{Goncalves1998}.

\subsection*{Appendix C: Standard Wavelet Spectra}\label{sec:wspctra1}
For data that scale, wavelet transforms have also been a widely used tool for estimating Hurst exponents. While several methods are available in the wavelet domain, most estimations rely on the properties of the wavelet spectrum. The wavelet spectrum can be described theoretically as follows. Considering the self-similar process given in (\ref{self-similar}), the detail wavelet coefficients satisfy

\begin{equation}\label{sf0}
 \displaystyle   d_{jk} \overset{d}{=} 2^{-j(H + 1/2)} d_{0k},
\end{equation}
for a fixed multiresolution level $j$ and under $L_2$ normalization. Also, if the  process in (\ref{self-similar}) is a zero-mean stochastic process and  possesses stationary increments (i.e., $\mathbb{E}(X) = 0$ and $X(t+h) - X(t)$ is independent of $t$), then $\mathbb{E}(d_{0k}) = 0$ and $\mathbb{E}(d_{0k}^2) = \mathbb{E}(d_{00}^2)$. As a result,

\begin{equation}\label{sf1}
    \displaystyle \mathbb{E}(d^2_{jk}) \propto 2^{-j(2H + 1)}.
\end{equation}
Equation (\ref{sf1}) provides a basis for estimating the Hurst exponent $H$ by taking logarithms of both sides.  The set of pairs $(j, S(j)) = (j,  log_2 \mathbb{E}(d^2_{jk})~)$, is called the wavelet spectrum. For the processes that scale, the  logarithms of average squared of detail wavelet coefficients (log energies) at different resolution scales exhibit a linear relationship with respect to the scale index, and this linear relationship, in particular the slope of the linear fit, is used to find an estimator for $H$. 

In practice, wavelet spectrum is computed as follows. Let  $\uw{d}_j = \{d_{1}, d_{2}, \cdots, d_{n}\}$ represents detail wavelet coefficients in the $j$th multiresolution level of wavelet transform of a signal $X$.  Then, the wavelet spectrum of $X$  is computed as
\begin{equation}\label{WT_spectra}
    S(j) = \log_2\left(\overline{\uw{d}_j^2}\right), ~J_0 \leq j \leq J-1,
\end{equation}
where $\overline{\uw{d}_j^2} (= \frac{1}{n}\sum_{i=1}^n d_i^2)$ is the average of squared wavelet coefficients (also known as log-energy) in level $j$, $J = \log_2(N)$, $N$ is the signal length, and $J_0$ ($1 \leq J_0 \leq J-1$) is the coarsest level of detail used to define the wavelet spectrum.

In order to estimate $H$, wavelet log-energies are regressed on the scale indexes. Based on the slope of the estimated regression model, the estimated value of $H$ is calculated as $H = -(slope +1)/2$.

\subsection*{Appendix D: Wavelet Entropy}\label{S4_Appendix}
Suppose wavelet decomposition of a signal $Y$ size $N \times 1$ and $\uw{d}_j = \{d_1, d_2, \cdots, d_n\}$ represents the set of detail wavelet coefficients at the resolution level $j$. Then, normalized Shannon wavelet entropy (WE) of $Y$ at $j$th scale level can be expressed as:

\begin{equation}\label{w-entropy}
    WE(j) = -\sum_{i = 1}^n \frac{d^2_i}{\sum_{i = 1}^n d^2_i}\log \frac{d^2_i}{\sum_{i = 1}^n d^2_i}.
\end{equation}

\subsection*{Appendix E: Multifractality}\label{sec:Multifractal1}
According to (Reidi et al., 1999) and (Reidi (2002), a multifractal spectrum is defined by a local singularity strength measure based on wavelets:

\begin{equation}\label{mfs1}
    \displaystyle \alpha(t) =  \lim_{k2^{-j} \to t} \left( -\frac{1}{j} \log_2|d_{j,k}|\right),
\end{equation}
where $d_{j,k}$ is the normalized wavelet coefficient corresponding to the basis that is $L_1$ normalized, that is $\phi_{jk}(t) = 2^j\phi(2^jt - k)$ and $\psi_{jk}(t) = 2^j\psi(2^jt - k)$ at scale $j$ and location $k$. 
Smaller values of $\alpha(t)$ correspond to larger oscillation in the signal and thus more singularity strengths. The frequency (in {\bf t}) of occurrence of a given singularity strength $\alpha$ is measured by the multifractal spectrum, denoted as $f(\alpha)$ analogously to Reidi (2002). That is, any inhomogeneous process has a set of local singularity measures and their distribution ($f(\alpha)$) forms the multifractal spectra.

A practical approach to compute multifractal spectrum makes use of the theory of large deviations, where $f$ would be interpreted as the rate function in a \emph{Large Deviation Principle}( $f$ measures how frequently (in {\bf k}) the observed $(-1/j)\log_2|d_{j,k}|$ deviate from the expected value for $\alpha$ in scales defined by $j$). More specifically, the multifractal spectrum depends on the concepts of \emph{partition function}, $T(q)$ and \emph{Legendre transform}.

The partition function, $T(q)$, is defined as

\begin{equation}\label{mfs2}
    \displaystyle T(q) =  \lim_{j \to \infty} \left( -\frac{1}{j} \log_2 \mathbb{E}(|d_{j,k}|^q) \right),
\end{equation}
$T(q)$ describes the limiting behavior of $q$th moment of a typical wavelet coefficient $d_{j,k}$ at the scale $j$ and location $k$. The moment order $q \in \mathbb{R}$ is within a certain range, covering negative numbers as well. Then, the multifractal formalism points that the multifractal spectrum can be calculated by taking the \emph{Legendre transform} of the log moment generating function (Riedi et al., 1999):

\begin{equation}\label{mfs3}
    \displaystyle f(\alpha) \simeq f_L(\alpha) := \inf_{q} [q\alpha - T(q)].
\end{equation}
It can be shown that $f(\alpha) = q\alpha - T(q)$ at $\alpha = T^{'}(q)$ given that $T^{''}(q) < 0$, and also $f_L(\alpha)$ converges to the true multifractal spectrum using the theory of large deviations (R.S. Ellis, (1984)).

To compute multifractal spectrum, we need a good estimator of the partition function. This estimator can be derived by rearranging (\ref{mfs2}) as follows.

\begin{equation}\label{mfs4}
    \mathbb{E} |d_{jk}|^q \sim 2^{-jT(q)} \quad \textrm{as} \quad j \to \infty.
\end{equation}
On the other hand, the $q$th moment of the wavelet coefficients of the power law process (Arneodo et al., 1998), holds the following expression:

\begin{equation}\label{mfs5}
    \mathbb{E}|d_{jk}|^q = C_q 2^{-jqH},
\end{equation}
where $H$ is the irregularity index (scaling exponent) and $C_q$ is a constant depending only on $q$. \\
Comparing (\ref{mfs4}) and (\ref{mfs5}), we can easily connect the partition function with the scaling exponent estimation procedure explained in section
\ref{sec:self-similar}. The use of linear regression has been a standard practice to identify $H$ since the values $\mathbb{E}(|d_{j,k}|^q))$ could be easily computed by moment matching method, estimating the partition function, $T(a)$ easy. We can express this more formally as  follows:

\begin{equation}\label{mfs6}
    \log_2 \widehat{S_j(q)} = -jT(q) + \epsilon_j,
\end{equation}
where $\displaystyle \widehat{S_j(q)} = \frac{1}{2^j}\sum_{k = 1}^{N2^{-j}}|d_{jk}|^q$ is the empirical $q$th moment of the wavelet coefficients, $N$ is the length of the signal, and $\epsilon$ is the error term introduced from the moment matching method when replacing the true moment with the empirical one. The ordinary least square is a convenient method for estimating the partition function.

After estimating function $T(q)$, the \emph{Legendre transform} is performed on the estimated the partition function. Since $\frac{\partial}{\partial q} (\alpha q - T(q)) = \alpha - T'(q)$ and $T''(q) < 0$, the maximum value of $\alpha q - T(q)$ is achieved at $a = T'^{(-1)}(\alpha)$. This indicates that performing Legendre transformation is divided into two steps: computing numerical derivative of  $T(q)$ using finite differences and then evaluating the value of Legendre spectrum at $\alpha = \widehat{T'(q)}$. It is important to note that the \emph{Legendre transform} cannot estimate the multifractal spectrum at arbitrary singularity strength $\alpha$. A set of multifractal spectrum values is determined by a range of $q$ values. The more $q$ values adopted, the smoother the multifractal spectrum. For example, a representative example multifractal spectrum can be found in Figure \ref{Mfrac}.

\end{document}